%% file: YS_arXiv2016_ABATLG_QHE.tex
\documentclass[aps,prl,twocolumn,superscriptaddress]{revtex4-1}
\usepackage{here}
\usepackage{graphicx}
\usepackage{subcaption}
\usepackage{braket}
\usepackage{color}
\usepackage{amsmath}
\begin{document}


\title{Landau level evolution driven by band hybridization in mirror symmetry broken ABA-stacked trilayer graphene}

\include{author}

\include{abstract}


\maketitle


\input{introduction}
\input{body}

\begin{acknowledgments}
Y. S. acknowledges support from JSPS KAKENHI Grant Number JP13J07387
and JSPS Program for Leading Graduate Schools (MERIT).
M. Y. acknowledges support from Canon Foundation and JSPS KAKENHI Grant Number JP25107003.
K. Watanabe and T. T. acknowledge support from JSPS KAKENHI Grant Number JP15K21722
and the Elemental Strategy Initiative conducted by the MEXT, Japan.
T. T. acknowledges support from JSPS Grant-in-Aid for Scientific Research A (No. 26248061) 
and JSPS Innovative Areas ``Nano Informatics'' (No. 25106006).
K. Wang and P. K. acknowledge support from ARO MURI (W911NF-14-1-0247).
S. T. acknowledges support from JSPS Grant-in-Aid for Scientific Research S (No. 26220710) and A (No. 16H02204).

Y. S. and T. Y. contributed equally to this work.
\end{acknowledgments}

\bibliography{QHE2016}

\end{document}

%% file: author.tex

\author{Y. Shimazaki}
\email{shimazaki@meso.t.u-tokyo.ac.jp}

\author{T. Yoshizawa}

\author{I. V. Borzenets}
\affiliation{Department of Applied Physics, University of Tokyo, Bunkyo-ku, Tokyo 113-8656, Japan}

\author{K. Wang}

\author{X. Liu}
\affiliation{Department of Physics, Harvard University, Cambridge, MA 02138, USA}

\author{K. Watanabe}

\author{T. Taniguchi}
\affiliation{National Institute for Materials Science, Tsukuba-shi, Ibaraki 305-0044, Japan}

\author{P. Kim}
\affiliation{Department of Physics, Harvard University, Cambridge, MA 02138, USA}
\affiliation{John A. Paulson School of Engineering and Applied Sciences, Harvard University, Cambridge, MA 02138, USA}

\author{M. Yamamoto}
\affiliation{Department of Applied Physics, University of Tokyo, Bunkyo-ku, Tokyo 113-8656, Japan}
\affiliation{PRESTO, JST, Kawaguchi-shi, Saitama 331-0012, Japan}

\author{S. Tarucha}
\affiliation{Department of Applied Physics, University of Tokyo, Bunkyo-ku, Tokyo 113-8656, Japan}
\affiliation{Center for Emergent Matter Science (CEMS), RIKEN, Wako-shi, Saitama 351-0198, Japan}


\date{\today}

%% file: abstract.tex
\begin{abstract}
Layer stacking and crystal lattice symmetry play important roles in the band structure and the Landau levels
of multilayer graphene. 
ABA-stacked trilayer graphene possesses mirror-symmetry-protected monolayer-like and bilayer-like band structures. 
Broken mirror symmetry by a perpendicular electric field therefore induces hybridization between these bands and 
various quantum Hall phases emerge.
We experimentally explore the evolution of Landau levels in ABA-stacked trilayer graphene under electric field.
We observe a variety of valley and orbital dependent Landau level evolutions.
These evolutions are qualitatively well explained by considering the hybridization between multiple
Landau levels possessing close Landau level indices and the hybridization
between every third Landau level orbitals due to the trigonal warping effect.
These observations are consistent with numerical calculations.
The combination of experimental and numerical analysis thus reveals the entire picture of Landau level evolutions
decomposed into the monolayer- and bilayer-like band contributions in ABA-stacked trilayer graphene.
\end{abstract}

%% file: introduction.tex
Recent progress on the study of quantum Hall effect (QHE) in graphene
revealed novel physics such as half-integer QHE in monolayer graphene
\cite{Zhang2005, Novoselov2005},
spin-valley quantum Hall ferromagnetism \cite{Young2012},
spin phase transition and quantum spin Hall state at charge neutrality point \cite{Maher2013, Young2014},
and Hofstadter's butterfly in graphene/h-BN superlattice \cite{Hunt2013, Ponomarenko2013, Dean2013}.

Graphene has a unique band structure and related properties drastically depending on layer number, stacking, and symmetry.
Monolayer graphene (MLG), AB-stacked bilayer graphene (BLG)
and ABC-stacked trilayer graphene (TLG) posess linear, parabolic, and cubic
band dispersions, respectively.
These systems have inversion symmetry and breaking inversion symmetry results in band gap formation.
For the case of multilayer graphene, the inversion symmetry of the system can be controlled
with perpendicular electric field \cite{Oostinga2007, Zhang2009, Shioya2012, Lui2011, Khodkov2015}.
In these multilayer systems, the integer and fractional quantum Hall phase transitions are also tunable with a perpendicular electric field
\cite{Weitz2010, Kim2011, Velasco2012, Maher2014, Young2014, Lee2014, Velasco2014, Lee2016, Shi2016}.

\begin{figure*}[t]
	\centering
	\includegraphics[width=7in]{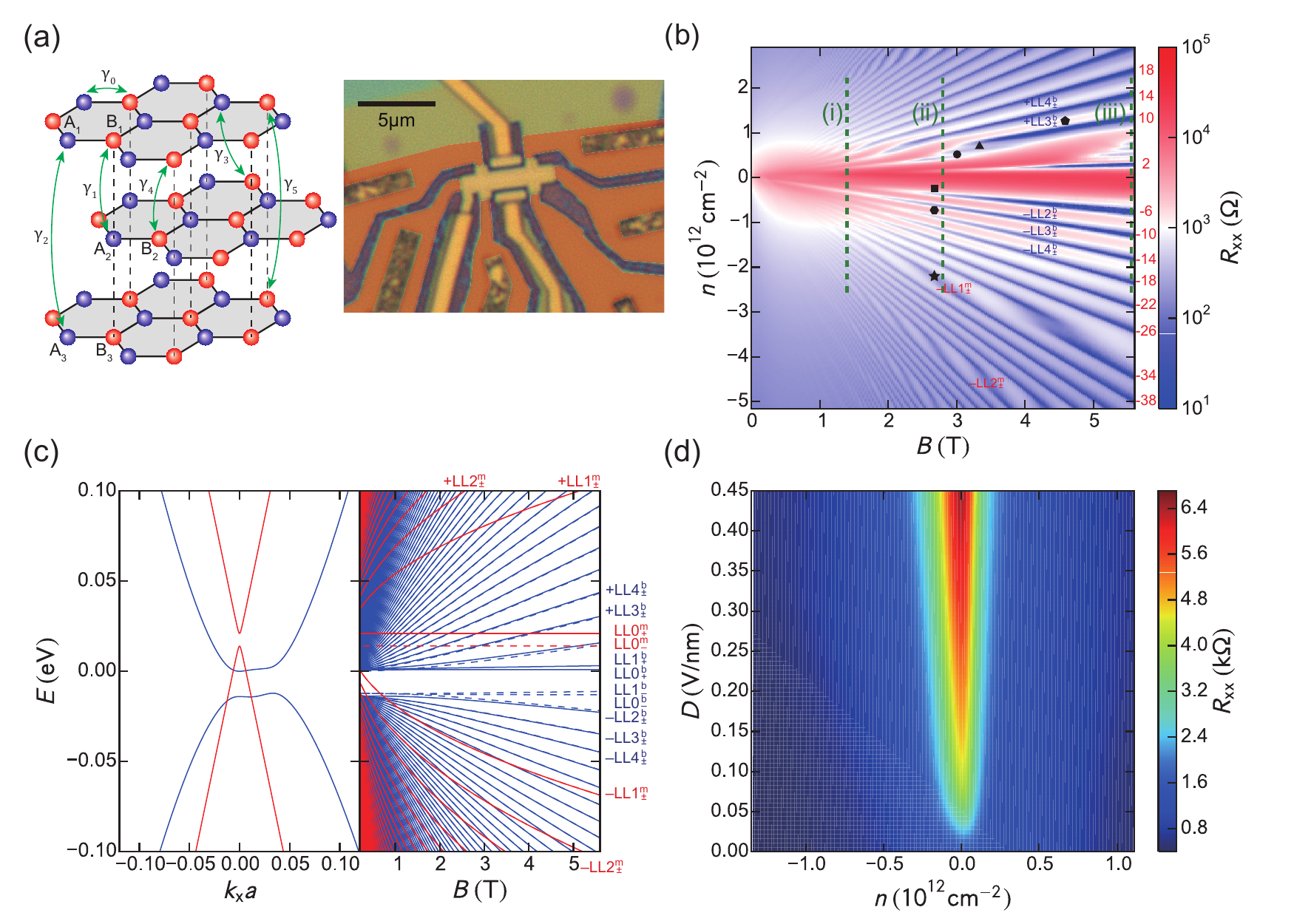}
	\caption{\raggedright 
	(a) Left : Schematic image of the ABA-stacked trilayer graphene with tight-binding
	parameters \cite{Koshino2009}. 
	    Right : Optical microscope image of the TLG device encapsulated between h-BN layers.
	    The device was shaped by reactive ion etching. The areas highlited in red are unetched and used as contact leads.
	(b) Carrier density $n$ and magnetic field $B$ dependence of longitudinal resistance $R_{xx}$ measured at $D = 0\ \rm{V/nm}$.
	The numbers in red on the right side of the figure show the filling factor 
	at each local minimum of $R_{xx}$.
	(c) Numerical calculation of a typical band structure of ABA-stacked TLG around K point (left figure) and
	    $B$ dependence of the Landau levels at $D = 0\ \rm{V/nm}$ (right figure).
	This calculation is performed using the tight-binding Hamiltonian with all tight-binding parameters
	\cite{Koshino2009, Koshino2011, Taychatanapat2011}.
	The band structure is plotted as a function of $k_x a$ for $k_y a = 0$
	where $(k_x, k_y)$ is the wave number vector measured from K point
	and $a = 0.246\ \mathrm{nm}$ is the lattice constant.
	The red (blue) colored lines show the MLG-like (BLG-like) bands at $B = 0\ \rm{T}$
	and their evolutions into the Landau levels.
	The solid (broken) lines in the right figure indicate the valley index $\rm K$ ($\rm K'$).
	(d) $n$ and $D$ dependence of $R_{xx}$ measured at 10 K.
	Gate capacitances are estimated so that the $R_{xx}$ peak lies along $n=0$.
	}
\end{figure*}

Trilayer graphene can have a different kind of stacking, called ABA-stacking (Fig. 1(a) left).
ABA-stacked TLG has mirror symmetry.
Reflecting this mirror symmetry, ABA-stacked TLG 
posesses a combined MLG-like and BLG-like band structure. 
The MLG-like band corresponds to mirror-symmetric wave function,
whereas the BLG-like band corresponds to mirror-antisymmetric wave function.
These two bands can hybridize when mirror symmetry is broken.
In previous works, the mirror symmetry of the system is controlled by a perpendicular electric field
\cite{Craciun2009, Henriksen2012, Zou2013, Lee2013}.
Therefore, ABA-stacked TLG is a novel platform to investigate the effect of
band hybridization on the QHE.
Under a perpendicular electric field,
ABA-stacked TLG is expected to show complex but rich Landau level (LL) evolution,
such as LL crossings and valley splitting due to the LL index dependent hybridization processes
\cite{Koshino2010, Morimoto2013, Serbyn2013}.
These featuers, however, have not yet been well revealed.
In gapped bilayer graphene, it has been shown that the trigonal warping effect results in three fold degenerate LLs at low magnetic field and small carrier density regime \cite{Varlet2014}.
In mirror symmetry broken ABA-stacked trilayer graphene, 
it is also expected that the trigonal warping affects LL orbital hybridization 
to induce anti-crossings between specific LLs \cite{Serbyn2013}.
At such anti-crossing points, the LL orbital states are tunable by an electric field; this may allow for further 
study on many-body phenomena such as quantum Hall ferromagnetism and fractional qunatum Hall effect.
Previous studies on the QHE in ABA-stacked TLG have revealed LL crossings
due to the existence of independent MLG-like and BLG-like bands
and unconventional sequence of the quantum Hall plateaus; 
which is explained by taking into account the all hopping parameters $\gamma_0$ to $\gamma_5$ 
(Fig. 1(a) left) and 
the onsite energy difference $\delta$ between A1, B2, A3 and B1, A2, B3
\cite{Koshino2011, Taychatanapat2011, Jhang2011, Henriksen2012, Lee2013}.
On the other hand, the perpendicular electric field dependence of the QHE in ABA-stacked TLG 
remained experimentally unclear, 
though the emergence or stabilization of a few novel quantum Hall plateaus have been observed
\cite{Henriksen2012, Lee2013}.

In this study, we reveal detailed picture of the evolution of 
the LL structure in ABA-stacked TLG under a perpendicular electric field.
Importantly, all tight-binding parameters, $\gamma_0$ to $\gamma_5$ and $\delta$, which describe the nature of the bulk graphite, play important roles in the LL structures of the ABA-stacked TLG as well as the onsite energy differences induced by an electric field.
By applying perpendicular electric field, we find series of LL crossings and anti-crossings which originate from the hybridized MLG-like and BLG-like bands
by breaking the mirror symmetry.
We also observe anti-crossing structures 
mediated by the trigonal warping effect, which hybridizes LLs with the LL index difference of integer times 3, for both electron and hole carriers.
These behaviors are well reproduced by numerical calculations based on the tight-binding model with the MLG-like and BLG-like LL orbital basis, involving all tight-binding parameters
$\gamma_0$ to $\gamma_5$, $\delta$ and average onsite energy differences
$\Delta_1$ between the layers induced by the electric field.

TLG is encapsulated between h-BN layers in order to 
achieve a high carrier mobility and to apply a higher electric field than that for
suspended devices (Fig. 1(a) right).
Both graphene and h-BN were obtained via mechanical exfoliation.
The number of graphene layers was confirmed optically via a contrast measurement.
As described below, the stacking of the TLG was confirmed as ABA from the measurement of the Landau fan diagram.
We used pickup technique to fabricate the BN/TLG/BN stack and made the Ohmic contacts to graphene 
from the edge \cite{Wang2013}.
To control the carrier density ($n$) and the perpendicular displacement 
electric field ($D$) independently, we adopt a dual-gate structure \cite{Oostinga2007}.
All measurements were performed at 1.7 K using VTI cryostat unless mentioned.

%% file: body.tex
\begin{table*}
\caption{\label{tab:table1}%
The correspondence between LL names and their wave function at $D = 0\ \rm{V/nm}$.
We adopted the notation of the LLs introduced by Ref. \cite{Serbyn2013}.
The numbering of the LLs is based on the dominant and larger LL orbital indices.
The superscript ``m'' (``b'') indicates that the LL is the eigenstate of
the monolayer (bilayer) block of the Hamiltonian.
The subscript ``$+$'' (``$-$'') indicates valley $\rm K$ ($\rm K'$).
$+{\rm LL}n^{\rm{m,b}}_{\pm}$ has larger energy than $-{\rm LL}n^{\rm{m,b}}_{\pm}$.
For the low energy, the wave function of bilayer block LLs are obtained from two band approximation of the original
$4 \times 4$ Hamiltonian, owing to the high energy of $\ket{\rm{A}_2}$ and $\frac{\ket{\rm{B}_1} + \ket{\rm{B}_3}}{\sqrt{2}}$.
Here we neglected the trigonal warping term ($\gamma_3$).
$a^{\rm m, b\pm}_{n}, b^{\rm m, b\pm}_{n}$ are non-zero coefficients.
}
\footnotesize
\begin{ruledtabular}
\begin{tabular}{rc|rc}
	LL & Wave function & LL & Wave function\\
	\hline
	$\rm{LL}0^{\rm{m}}_{+}$ & $\ket{0} \otimes \frac{\ket{\rm{B}_1} - \ket{\rm{B}_3}}{\sqrt{2}} \otimes \ket{\rm{K}}$
	& $\rm{LL}0^{\rm{m}}_{-}$ & $\ket{0} \otimes \frac{\ket{\rm{A}_1} - \ket{\rm{A}_3}}{\sqrt{2}} \otimes \ket{\rm{K'}}$\\
	$\rm{LL}0^{\rm{b}}_{+}$ & $\ket{0} \otimes \ket{\rm{B}_2} \otimes \ket{\rm{K}}$
	& $\rm{LL}0^{\rm{b}}_{-}$ & $\ket{0} \otimes \frac{\ket{\rm{A}_1} + \ket{\rm{A}_3}}{\sqrt{2}} \otimes \ket{\rm{K'}}$ \\
	$\rm{LL}1^{\rm{b}}_{+}$ & $\ket{1} \otimes \ket{\rm{B}_2} \otimes \ket{\rm{K}}$
	& $\rm{LL}1^{\rm{b}}_{-}$ & $\ket{1} \otimes \frac{\ket{\rm{A}_1} + \ket{\rm{A}_3}}{\sqrt{2}} \otimes \ket{\rm{K'}}$ \\
	$+{\rm LL}n^{\rm{m}}_{+}$ & $\left(a^{\rm{m}+}_{n}\ket{n-1} \otimes \frac{\ket{\rm{A}_1} - \ket{\rm{A}_3}}{\sqrt{2}} + b^{\rm{m}+}_{n}\ket{n} \otimes \frac{\ket{\rm{B}_1} - \ket{\rm{B}_3}}{\sqrt{2}}\right) \otimes \ket{\rm{K}}$
	&$+{\rm LL}n^{\rm{m}}_{-}$ & $\left(a^{\rm{m}-}_{n}\ket{n} \otimes \frac{\ket{\rm{A}_1} - \ket{\rm{A}_3}}{\sqrt{2}} + b^{\rm{m}-}_{n}\ket{n-1} \otimes \frac{\ket{\rm{B}_1} - \ket{\rm{B}_3}}{\sqrt{2}}\right) \otimes \ket{\rm{K'}}$\\
	$-{\rm LL}n^{\rm{m}}_{+}$ & $\left((b^{\rm{m}+}_{n})^*\ket{n-1} \otimes \frac{\ket{\rm{A}_1} - \ket{\rm{A}_3}}{\sqrt{2}} - (a^{\rm{m}+}_{n})^*\ket{n} \otimes \frac{\ket{\rm{B}_1} - \ket{\rm{B}_3}}{\sqrt{2}}\right) \otimes \ket{\rm{K}}$
	&$-{\rm LL}n^{\rm{m}}_{-}$ & $\left((b^{\rm{m}-}_{n})^*\ket{n} \otimes \frac{\ket{\rm{A}_1} - \ket{\rm{A}_3}}{\sqrt{2}} - (a^{\rm{m}-}_{n})^*\ket{n-1} \otimes \frac{\ket{\rm{B}_1} - \ket{\rm{B}_3}}{\sqrt{2}}\right) \otimes \ket{\rm{K'}}$ \\
	$+{\rm LL}n^{\rm{b}}_{+}$ & $\left(a^{\rm{b}+}_{n}\ket{n-2} \otimes \frac{\ket{\rm{A}_1} + \ket{\rm{A}_3}}{\sqrt{2}} + b^{\rm{b}+}_{n}\ket{n} \otimes \ket{\rm{B}_2}\right) \otimes \ket{\rm{K}}$
	&$+{\rm LL}n^{\rm{b}}_{-}$ & $\left(a^{\rm{b}-}_{n}\ket{n} \otimes \frac{\ket{\rm{A}_1} + \ket{\rm{A}_3}}{\sqrt{2}} + b^{\rm{b}-}_{n}\ket{n-2} \otimes \ket{\rm{B}_2}\right) \otimes \ket{\rm{K'}}$ \\
	$-{\rm LL}n^{\rm{b}}_{+}$ & $\left((b^{\rm{b}+}_{n})^*\ket{n-2} \otimes \frac{\ket{\rm{A}_1} + \ket{\rm{A}_3}}{\sqrt{2}} - (a^{\rm{b}+}_{n})^*\ket{n} \otimes \ket{\rm{B}_2}\right) \otimes \ket{\rm{K}}$
	&$-{\rm LL}n^{\rm{b}}_{-}$ & $\left((b^{\rm{b}-}_{n})^*\ket{n} \otimes \frac{\ket{\rm{A}_1} + \ket{\rm{A}_3}}{\sqrt{2}} - (a^{\rm{b}-}_{n})^*\ket{n-2} \otimes \ket{\rm{B}_2}\right) \otimes \ket{\rm{K'}}$
\end{tabular}
\end{ruledtabular}
\normalsize
\end{table*}

\begin{table}[b]
\caption{\label{tab:table2}%
Analytically derived energies for each LLs at $D = 0\ \rm{V/nm}$
by neglecting the trigonal warping term ($\gamma_3$).
$\zeta$ is defined by $\zeta=3eBa^2{\gamma_0}^2/(4\hbar{\gamma_1}^2)$, where $a$ is lattice constant.
}
\begin{ruledtabular}
\begin{tabular}{rc}
	LL & Energy\\
	\hline
	$\rm{LL}0^{\rm{m}}_{+}$ & $\delta - \frac{1}{2}\gamma_5 + \Delta_2$\\
	$\rm{LL}0^{\rm{m}}_{-}$ & $- \frac{1}{2}\gamma_2 + \Delta_2$\\
	$\rm{LL}0^{\rm{b}}_{+}$ & $- 2\Delta_2$\\
	$\rm{LL}1^{\rm{b}}_{+}$ & $- 2\Delta_2 + \zeta \left( \frac{\gamma_5}{2} + \delta + \Delta_2 \right)$\\
	$\rm{LL}0^{\rm{b}}_{-}$ & $ \frac{\gamma_2}{2}+\Delta_2$\\
	$\rm{LL}1^{\rm{b}}_{-}$ & $ \frac{\gamma_2}{2}+\Delta_2 + \zeta(\delta - 2\Delta_2)$
\end{tabular}
\end{ruledtabular}
\end{table}

\begin{figure*}[t]
	\centering
	\includegraphics[width=7in]{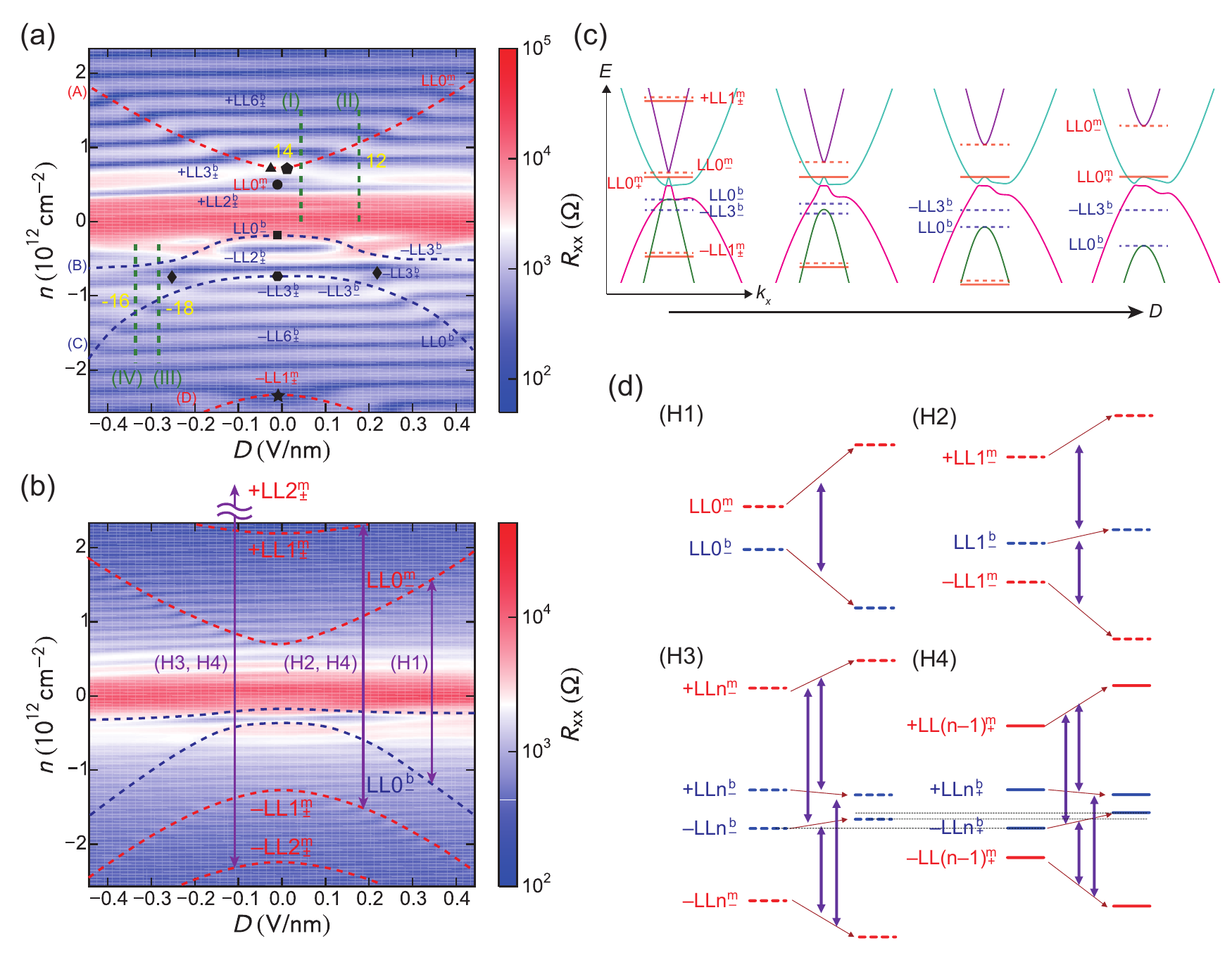}
	\caption{\raggedright 
	(a) and (b) $n$ and $D$ the dependences of $R_{xx}$ under $B = 2.8\ \rm{T}$, and $1.4\ \rm{T}$, respectively.
	The broken curves in red (blue) are eye guides, which 
	follow the evolution of the specific LLs in the MLG-like (BLG-like) bands.
	The marks in black indicate the LLs, corresponding to the ones shown in Fig. 1(b)
	except for the LL marked by the diamonds.
	The numbers in yellow indicate the filling factors between the LLs.
	The purple double sided arrows indicate the LLs involved in each hybridization process.
	(c) Calculated band structure evolution with $D$, The LLs are schematically shown
	by the solid (broken) lines, which exist at valley $\rm K$ ($\rm K'$).
	The following tight-binding parameters were used for the calculation of the band structure evolution : 
	$\gamma_0 = 3.1\ \mathrm{eV},
	\gamma_1 = 0.39\ \mathrm{eV},
	\gamma_2 = -0.028\ \mathrm{eV},
	\gamma_3 = 0.315\ \mathrm{eV},
	\gamma_4 = 0.041\ \mathrm{eV},
	\gamma_5 = 0.05\ \mathrm{eV}$,
	$\delta = 0.034\ \mathrm{eV}$,
	and 
	$\Delta_2 = 0\ \mathrm{eV}$,
	.
	$\Delta_1$ was modulated from $0\ \mathrm{eV}$ to $0.06\ \mathrm{eV}$
	at regular intervals.
	(d) Schematic image of the hybridization processes with $D$. The purple double sided arrows indicate
	pairs of LLs which directly hybridize under $D$.
	}
\end{figure*}

\begin{figure*}[t]
	\centering
	\includegraphics[width=7in]{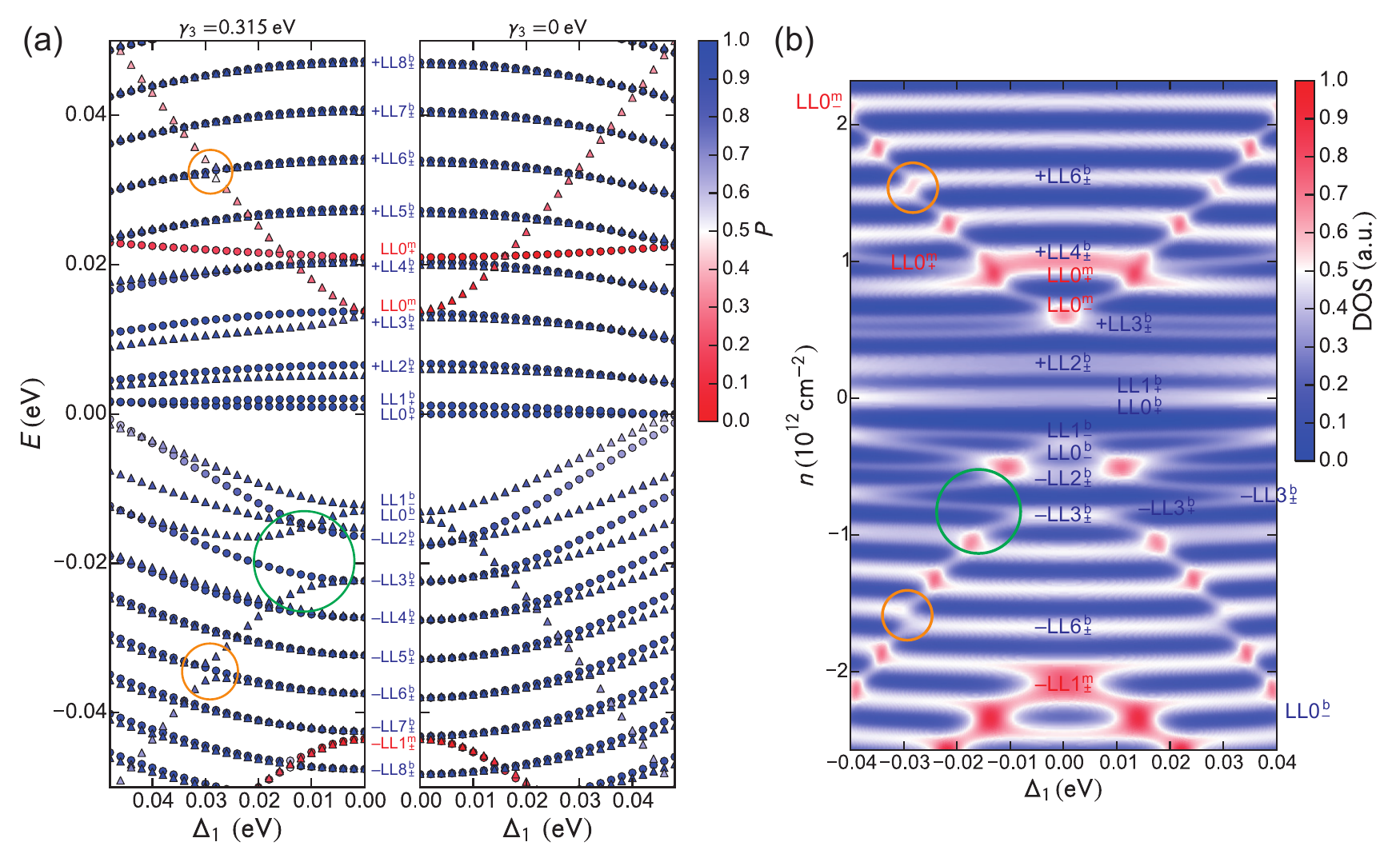}
	\caption{\raggedright 
	(a) Numerically calculated $\Delta_1$ dependence of the LLs energies at $B = 2.8\ \rm{T}$
	from the tight-binding model. 
	The parameters asuumed are the same as used in the calculation of Fig. 1(c) in the left graph,
	and the same except for $\gamma_3 = 0\ \rm{eV}$ in the right graph.
	The colorscale indicates the portion of the wave function belongings:
	the wave function of the blue (red) points mainly belongs to the bilayer (monolayer) block.
	The circle (triangle) shaped points belong to $\rm{K}$ ($\rm{K'}$) valley.
	The green and orange circles indicate the anti-crossings between LLs due to the trigonal warping.
	(b) Numerically calculated $n$ and $\Delta_1$ dependences of the LLs DOS at $B = 2.8\ \rm{T}$ 
	from the tight-binding model. The parameters are the same as used in the calculation of Fig. 1(c).
	The green and orange circles indicate the anti-crossings between LLs due to the trigonal warping.
	}
\end{figure*}

The Landau fan diagram at zero $D$ is shown in Fig. 1(b).
The top and back gate voltages are converted to $n$ and $D$ 
using the estimated top and back gate capacitances (see Fig. 1(d)).
We observe a number of LL crossings and asymmetry between the electron side and the hole side.
These features have been reported in ABA-stacked TLG \cite{Taychatanapat2011, Lee2013}, and
are accounted for by the coexistence of MLG-like and BLG-like bands, which are gapped and energetically detuned
from each other owing to the tight-binding parameters $\gamma_2$, $\gamma_5$, $\delta$, and $\Delta_2=\frac{1}{6}[(U_1-U_2)-(U_2-U_3)]$ defined by the onsite energy of each layer, $U_1$, $U_2$, and $U_3$.

We numerically calculated the band structure around K point
(left graph) and $B$ dependence of the LLs energies
at $D = 0\ \mathrm{V/nm}$ (right graph) for comparison as shown in Fig. 1(c).
The calculation was performed using the tight-binding model \cite{Koshino2009, Koshino2011}
with the parameters, 
$\gamma_0 = 3.1\ \mathrm{eV},
\gamma_1 = 0.39\ \mathrm{eV},
\gamma_2 = -0.028\ \mathrm{eV},
\gamma_3 = 0.315\ \mathrm{eV},
\gamma_4 = 0.041\ \mathrm{eV},
\gamma_5 = 0.05\ \mathrm{eV}$
and
$\delta = 0.046\ \mathrm{eV}$
 adopted from Ref. \cite{Taychatanapat2011}.
Approximated wave functions of the LLs for the low energy at $D = 0\ \rm{V/nm}$ are summarized in Table \ref{tab:table1}.
We define labeling of the LLs in the mirror-symmetric monolayer (m) and mirror-antisymmetric bilayer (b) by LL orbital basis.
(Note $\frac{\ket{\rm A_1} - \ket{\rm A_3}}{\sqrt{2}}$ is mirror-symmetric owing to the antisymmetric property 
of $\pi$ orbitals.)
Their energies, derived analytically by neglecting the small correction by $\gamma_3$, are shown in Table \ref{tab:table2} for the small LL indices.
The LL labels are shown visually in Fig. 1(c) (right).

By comparing Fig. 1(b) with Fig. 1(c), 
the LLs marked by the pentagon, hexagon, and stars in Fig. 1(b) are identified as
$+{\rm LL}3^{\rm{b}}_{\pm}$, $-{\rm LL}3^{\rm{b}}_{\pm}$ and $-{\rm LL}1^{\rm{m}}_{\pm}$, respectively.
The LLs marked by the triangle and circle in Fig. 1(b)
are identified as either LL$0^{\rm{m}}_{+}$ or LL$0^{\rm{m}}_{-}$.
The LL indicated by the square mark in Fig. 1(b) is identified
as either LL$0^{\rm{b}}_{\pm}$ or LL$1^{\rm{b}}_{\pm}$.
Depending on the tight-binding parameters such as $\gamma_2, \gamma_5, \delta$, and $\Delta_2$,
the order of the LL energies with small indices changes (see Table \ref{tab:table2})\cite{Serbyn2013}.
Therefore, we cannot uniquely identify the indices for the LLs indicated by the triangle, circle and square marks.

We then investigated the $D$ dependence of the transport characteristics. $D$ modifies the band structure and induces hybridization between the MLG-like and BLG-like bands.
Fig. 1(d) shows the $n$ and $D$ dependences of the longitudinal resistance $R_{xx}$ measured at 10K at $B = 0 \rm{T}$.
The $n$ and $D$ values were calculated using the top and back gate capacitances, which were estimated from the slope of
the $R_{xx}$ peak position in the plane defined by the top and back gate voltages and the known parameters of the gate insulators: 
The thickness, and dielectric constant of $\rm SiO_2$ is 285nm, and 3.9, respectively,
and the thickness of the top, and bottom h-BN layers is 44.1nm, and 52.7nm, respectively.
The $R_{xx}$ peak at $n = 0$ increases as $D$ increases.
This is understood by the reduction of the density of states around the charge neutrality point (CNP)
by the band hybridization due to the mirror symmetry breaking under $D$ \cite{Koshino2009, Henriksen2012}.
Such band evolution with $D$, expected from the tight-binding model with full parameters \cite{Koshino2009, Serbyn2013}, is shown in Fig. 2(c).


Figs. 2(a), and 2(b) show the $n$ and $D$ dependences of $R_{xx}$ for $B = 2.8\ \rm{T}$, and $1.4\ \rm{T}$, respectively.
The $R_{xx}$ measured along $D = 0\ \rm{V/nm}$ in Figs. 2(a), and 2(b) 
corresponds to the $R_{xx}$ measured along the broken green lines (i), and (ii) in Fig. 1(c), respectively.
In Figs. 2(a) and 2(b), we see a number of LLs parallel to the $D$ axis.
By comparing them with Fig. 1(c) at $D = 0\ \rm{V/nm}$,
most of these LLs, such as the LL marked by the pentagon ($+{\rm LL}3^{\rm{b}}_{\pm}$), 
are found to originate from the BLG-like bands.
In addition, several sequences of LL crossings identified by the kinks of the parallel LLs are observed in Figs. 2(a) and 2(b).
We put eye guides over some of the LLs which show characteristic evolution with $D$ leading to LL crossings.
As shown in Fig. 2(c), with increasing $D$, the MLG-like bands
move away from each other while the BLG-like bands gradually 
overlap due to the band hybridization.
It is therefore expected that 
most of the LLs in the MLG-like bands move away from the charge neutrality point (CNP) while
most of the LLs in the BLG-like bands gradually approach the CNP.

These trends are clearly observed in Fig. 2(b).
By following the LLs indicated by the eye guides shown in red
and comparing them with those in Fig. 1(b) at $D = 0\ \rm{V/nm}$,
it is found that these LLs originate from the MLG-like bands. 
One of the LLs indicated by the eye guides shown in blue
quickly moves away from the CNP with $D$.
However, in comparison with Fig. 1(b), 
this LL is found to originate from the BLG-like band.

The detailed evolution of LLs under $D$ is understood by considering
the sublattice degree of freedom
of LL wave functions summarized in Table \ref{tab:table1} and their hybridization.
Note that $\ket{n}$ is the LL orbital with index $n$.
$D$ affects $\Delta_1$, which is half of the energy detuning between the top and bottom layers.
Under $D$, $\frac{\ket{\rm{A}_1} - \ket{\rm{A}_3}}{\sqrt{2}}$ (monolayer) 
and $\frac{\ket{\rm{A}_1} + \ket{\rm{A}_3}}{\sqrt{2}}$ (bilayer) of the same $\ket{n}$ and valley hybridize.
From Table \ref{tab:table1}, we can identify four kinds of such hybridization processes:\\ 
(H1) $\rm{LL}0^{\rm{m}}_{-}$ and $\rm{LL}0^{\rm{b}}_{-}$\\
(H2) $\rm{LL}1^{\rm{b}}_{-}$ and $\pm\rm{LL}1^{\rm{m}}_{-}$\\
(H3) $\pm{\rm LL}n^{\rm{m}}_{-}$ and $\pm{\rm LL}n^{\rm{b}}_{-}$ ($n \geq 2$) \\
(H4) $\pm{\rm LL}(n-1)^{\rm{m}}_{+}$ and $\pm{\rm LL}n^{\rm{b}}_{+}$ ($n \geq 2$).\\
We schematically show these processes in Fig. 2(d).
In (H1), two LLs, $\rm{LL}0^{\rm{m}}_{-}$ and $\rm{LL}0^{\rm{b}}_{-}$, move away from each other.
In the rest of (H2), (H3), and (H4), two LLs in the MLG-like bands
and one or two LLs in the BLG-like bands hybridize.
For not too small $B$, the LLs in the BLG-like bands usually stay between
the two LLs in the MLG-like bands.
Therefore, the two LLs in the MLG-like bands move away from each other for increasing $D$.
On the other hand, the LLs in the BLG-like bands are repelled by them from both sides and 
slightly get closer to each other.

Note in the $D$ range studied here, hybridization between $\frac{\ket{\rm B_1}-\ket{\rm B_3}}{\sqrt{2}}$
and $\frac{\ket{\rm B_1}+\ket{\rm B_3}}{\sqrt{2}}$ has a minor influence
on the LLs owing to the large energy gap between them.
Therefore the energy of $\rm{LL}0^{\rm{m}}_{+}$ is almost unchanged with $D$.
$\rm{LL}0^{\rm{b}}_{+}$ and $\rm{LL}1^{\rm{b}}_{+}$, which are polarized to $\ket{\rm B_2}$,
are also expected to be almost independent of $D$.

In Fig. 2(c), we schematically show the expected evolution of the specific LLs with $D$.
The top of the valence band of the BLG-like bands
splits off into the top of the green-colored band and that of the magenta-colored band owing to the hybridization
with the valence band of the MLG-like bands.
One of the hybridized bands then moves away towards the lower energy
and forms a new valence band (green).
This process leads to the observation that $\rm{LL}0^{\rm{b}}_{-}$ moves away 
with other LLs of the MLG-like bands with increasing $D$,
since only the top of the green-colored band has the BLG-like $\frac{\ket{\rm{A}_1} + \ket{\rm{A}_3}}{\sqrt{2}}$
component and the rest originates from the MLG-like components.
On the other hand, the top of the valence band of the MLG-like bands
(top of the band colored with light blue) remains unchanged with $D$ although there is hybridization with the BLG-like band.
This leads to the observation in Fig. 2(a) that the energy of $\rm{LL}0^{\rm{m}}_{+}$ as well as those of LLs in the BLG-like bands remains almost constant.

\begin{figure*}[t]
	\centering
	\includegraphics[width=7in]{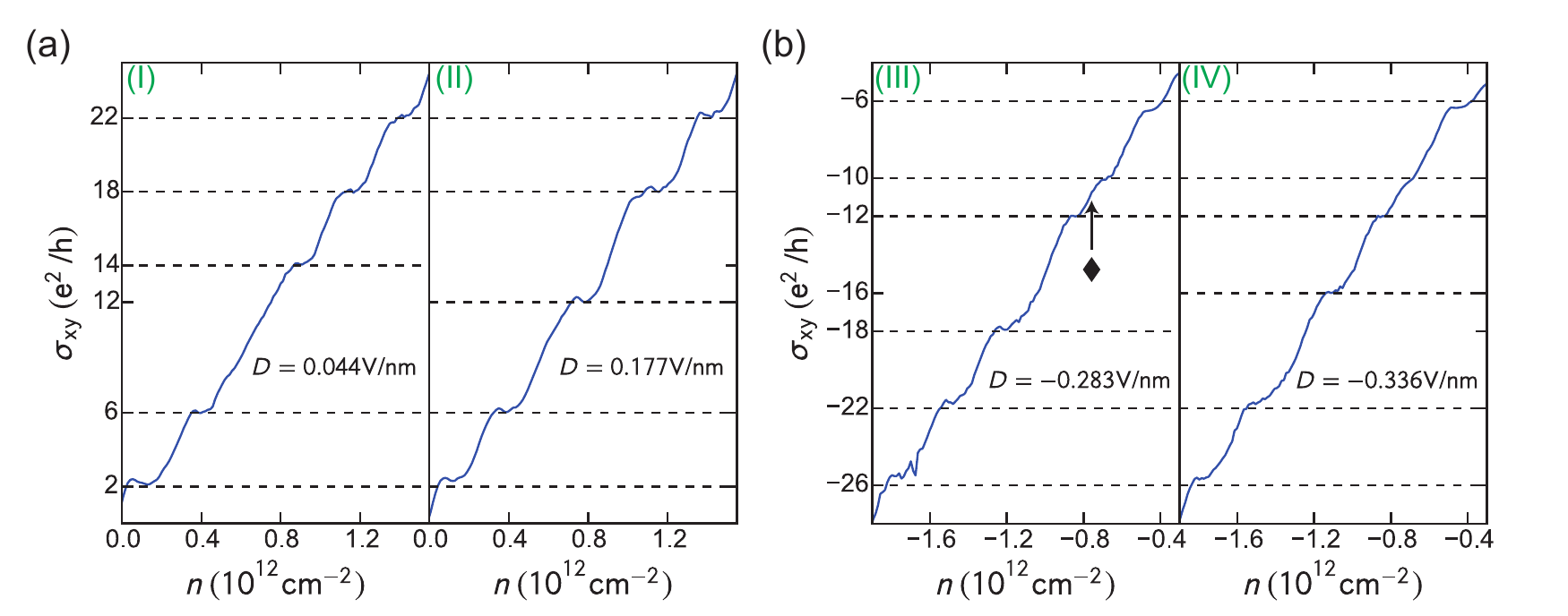}
	\caption{\raggedright (a) $n$ dependence of $\sigma_{xy}$ for each fixed $D$ at $B=2.8\ \rm{T}$: 
	$D = 0.044\ \rm{V/nm}$ (I) and $D = 0.177\ \rm{V/nm}$ (II).
	(b) $n$ dependence of $\sigma_{xy}$ for each fixed $D$ at $B=2.8\ \rm{T}$: 
	$D = -0.283\ \rm{V/nm}$ (III) and $D = -0.336\ \rm{V/nm}$ (IV).
	(I) to (IV) indicate $\sigma_{xy}$ simultaneously measured with $R_{xx}$ along the broken green lines of (I) to (IV) in Fig. 2(a).
	}
\end{figure*}

From the above considerations, we identified that the LL marked by the triangle, circle, and square in Fig. 2(a)
are $\rm{LL}0^{\rm{m}}_{-}$, $\rm{LL}0^{\rm{m}}_{+}$, and $\rm{LL}0^{\rm{b}}_{-}$, respectively.
The hybridization process (H1) separates $\rm{LL}0^{\rm{m}}_{-}$ and $\rm{LL}0^{\rm{b}}_{-}$ with increasing $D$ while $\rm{LL}0^{\rm{m}}_{+}$ stays at the constant energy owing to the weak influence of the $\frac{\ket{\rm{B}_1} + \ket{\rm{B}_3}}{\sqrt{2}}$ orbitals. Note that evolutions of these LLs are $B$ independent (common in Figs. 2(a) and 2(b)) around $D=0$. This further supports that these are the zeroth LLs, representing the edges of the bands as described in Fig. 2(c). 

However, in Fig. 2(a) the eye guide (B) passes through $\rm{LL}0^{\rm{b}}_{-}$ at $D = 0\ \rm{V/nm}$, 
and becomes almost independent of $D$ for $|D| > 0.2\ {\rm V/nm}$.
In contrast, the eye guide (C), which passes through $-\rm{LL}3^{\rm{b}}_{\pm}$ at $D = 0\ \rm{V/nm}$, 
shifts away from $n \sim -0.7\ \rm{cm^{-2}}$ for $|D| > 0.2\ {\rm V/nm}$.
The observed behavior around $|D| \sim 0.2\ {\rm V/nm}$ is interpreted
as the anti-crossing between $\rm{LL}0^{\rm{b}}_{-}$ and $-\rm{LL}3^{\rm{b}}_{-}$.
It has been pointed out that the trigonal warping term $\gamma_3$
hybridizes every third LLs of the BLG-like bands \cite{Serbyn2013}.
This means that the LLs of the BLG-like bands which mainly consist of $\ket{n}$,
are influenced by $\ket{n\pm 3}, \ket{n\pm 6}, \cdots$.
Therefore, when we consider the evolution of $\rm{LL}0^{\rm{b}}_{-}$ and $\rm{LL}0^{\rm{m}}_{-}$ by the hybridization under $D$,
$\pm\rm{LL}3^{\rm{m, b}}_{-}$, $\pm\rm{LL}6^{\rm{m, b}}_{-}$, $\cdots$
also have to be taken into account.

In contrast to $-\rm{LL}3^{\rm{b}}_{-}$, which shows anti-crossing with $\rm{LL}0^{\rm{b}}_{-}$,
$-\rm{LL}3^{\rm{b}}_{+}$ remains unaffected.
Therefore, $-\rm{LL}3^{\rm{b}}_{+}$, marked by the diamonds in Fig. 2(a),
splits off from $-\rm{LL}3^{\rm{b}}_{-}$, crosses over the anti-crossing structure,
and merges again with $-\rm{LL}3^{\rm{b}}_{-}$.
Note this is the first observation of the trigonal warping mediated valley splitting.

\begin{figure*}[t]
	\centering
	\includegraphics[width=7in]{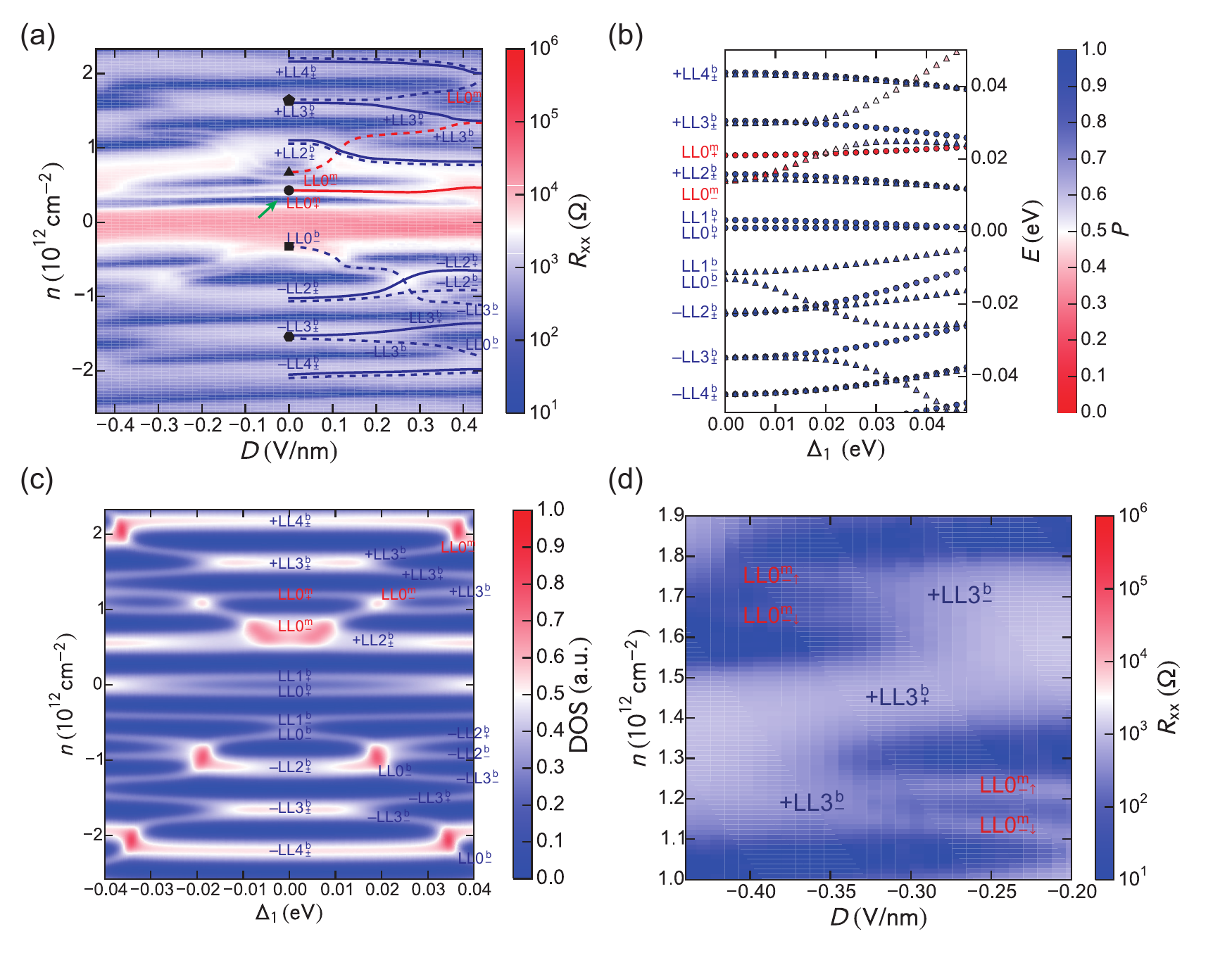}
	\caption{\raggedright 
	(a) $n$ and $D$ dependences of $R_{xx}$ measured at $B = 5.6\ \rm{T}$.
	The curves in red (blue) are eye guides, which 
	follow the evolution of the specific LLs in the MLG-like (BLG-like) bands.
	The LLs indicated by the solid (broken) curves belong to $\rm{K}$ ($\rm{K'}$) valley.
	(b) Numerically calculated $\Delta_1$ dependence of the LLs energies at $B = 5.6\ \rm{T}$.
	The parameters assumed are the same as used in the calculation of Fig. 1(c).
	The colorscale indicates the portion of the wave function belongings:
	the wave function of the blue (red) points mainly belongs to the bilayer (monolayer) block.
	The circle (triangle) shaped points belong to $\rm{K}$ ($\rm{K'}$) valley.
	(c) Numerically calculated $n$ and $\Delta_1$ dependences of the LLs DOS at $B = 5.6\ \rm{T}$.
	The parameters assumed are the same as used in the calculation of Fig. 1(c).
	(d) Extracted and magnified figure from the area around the anti-crossing 
	between $\rm{LL}0^{\rm{m}}_{-}$ and $\rm{LL}0^{\rm{b}}_{-}$.
	}
\end{figure*}

To further confirm this scenario, we show numerical calculations of LLs energies $E_{\rm LL}$ vs $\Delta_1$
at $B = 2.8\ \rm{T}$,
for $\gamma_3 = 0.315\ \rm{eV}$ (left graph) and $\gamma_3 = 0\ \rm{eV}$ (right graph) in Fig. 3(a),
where $\Delta_1$ is half the energy detuning between the top and bottom layers, which is controlled by $D$.
The calculation was based on Ref. \cite{Serbyn2013}.
Here, we used the same tight-binding parameters 
as used for the calculation of Fig. 1(c).
The experimental result (Fig. 2(a)) including the anti-crossing described above is well reproduced in the left graph of Fig. 3(a).
We clearly see the anti-crossing structure between $\rm{LL}0^{\rm{b}}_{-}$ and $-\rm{LL}3^{\rm{b}}_{-}$ for $\gamma_3 = 0.315\ \rm{eV}$ (indicated by the green circle),
but not for $\gamma_3 = 0\ \rm{eV}$.
Though, we see small anti-crossing structures between $\rm{LL}0^{\rm{b}}_{-}$ and $-\rm{LL}6^{\rm{b}}_{-}$, $\rm{LL}0^{\rm{m}}_{-}$ and $+\rm{LL}6^{\rm{b}}_{-}$
for $\gamma_3 = 0.315\ \rm{eV}$ in the calculation (indicated by the orange circles),
they are not visible in Fig. 2(a).
This is attributed to line broadening of the LLs.
In Fig. 3(b), we show the $\Delta_1$ and $n$ dependences of the LLs density of states (DOS)
 calculated from the energy level $E_{\rm LL}$ in Fig. 3(a) (left graph). 
 Here the DOS of each LL is assumed as:
${\rm DOS}(E;E_{\rm LL}) = \frac{2B}{h/e}\frac{1}{\pi}\frac{\Gamma/2}{(E-E_{\rm LL})^2 + (\Gamma/2)^2}$ 
using the level broadening $\Gamma = 2\ {\rm meV}$,
where $h$, and $e$ are the Planck constant, and elementary charge, respectively \cite{Taychatanapat2011}.
We find good agreement of this calculation with the experimental result in Fig. 2(a).
Note that the order of some LLs in Figs. 3(a) and 3(b) is different from that in Fig. 2(a).
For example, the order of $\rm{LL}0^{\rm{m}}_{+}$ and $\rm{LL}0^{\rm{m}}_{-}$, which is determined by $\gamma_2, \gamma_5$ and $\delta$ (see Table \ref{tab:table2}), is opposite.
This is probably due to the environmental difference 
or the imperfection of the estimation of the tight binding parameters 
in previous work \cite{Taychatanapat2011} in which $\Delta_2$ is neglected.

The consistency of the LLs assignment between the experiment and calculation is further confirmed 
by investigating the degeneracy of the observed LLs.
The LLs indicated by the eye guides (A) and (C) ($\rm{LL}0^{\rm{m}}_{-}$ and $-\rm{LL}3^{\rm{b}}_{-}$) in Fig. 2(a)
are considered to be valley polarized to ${\rm K}'$.
Therefore, the filling factor should change by two across these lines.
Fig. 4 shows the $n$ dependence of the Hall conductivity $\sigma_{xy}$ at different $D$ simultaneously measured with $R_{xx}$ in Fig. 2(a).
(I) to (IV) indicate the correspondence between the green broken lines (I) to (IV) in Fig. 2(a).
In Fig. 4(a), the filling factor around $n\sim 0.8 \times 10^{12}\ {\rm cm^{-2}}$
changes from 14 (left graph) to 12 (right graph).
Also, in Fig. 4(b), the filling factor around $n\sim -1.2 \times 10^{12}\ {\rm cm^{-2}}$
changes from -18 (left graph) to -16 (right graph).
These observations are consistent with the interpretation that the evolving LLs indicated by
the eye guides (A) and (C) are valley polarized.
In the left graph of Fig. 4(b), we see the change of the filling factor
from -12 to -10 around $n\sim -0.8\times10^{12}\rm{cm}^{-2}$ across the LL 
marked by the diamonds in Fig. 2(a).
Therefore, this LL has two-fold degeneracy.
This is also consistent with the interpretation that 
the anti-crossing between $\rm{LL}0^{\rm{m}}_{-}$ and $-\rm{LL}3^{\rm{b}}_{-}$ results in
the valley splitting of $-\rm{LL}3^{\rm{b}}_{\pm}$ and the LL marked by
the diamonds is $-\rm{LL}3^{\rm{b}}_{+}$.

Finally, we show the LLs evolution with $n$ and $D$ at a high $B$ of $5.6\ {\rm T}$ in Fig. 5(a).
The $R_{xx}$ measured along $D = 0\ \rm{V/nm}$ in Fig. 5(a) 
corresponds to the $R_{xx}$ measured along the broken green line (iii) in Fig. 1(c).
At this $B$, many fine structures are visible,
for instance spin splittings of LLs for $\rm{LL}0^{\rm{m}}_{\pm}$ and $\rm{LL}0^{\rm{b}}_{-}$.
Fig. 5(b) shows numerically calculated $\Delta_1$ dependence of the LLs energies at $B = 5.6\ \rm{T}$,
using the same tight-binding parameters as in the calculation of Fig. 3(a).
In Fig. 5(c), we show the $\Delta_1$ and $n$ dependences of the LLs DOS
calculated in the same manner as used for the calculation of Fig. 3(b).
The LLs evolution in Fig. 5(a) compares well to the calculation of Figs. 5(b) and 5(c).
From this comparison, we identify the LLs evolution as indicated by the eye guides in Fig. 5(a).

Around $D \sim 0.1\ \rm{V/nm}$, we see a kink structure in the evolution of $\rm{LL}0^{\rm{b}}_{-}$ 
indicating the presence of an additional LL crossing. 
$\rm{LL}0^{\rm{b}}_{+}$ and $\rm{LL}1^{\rm{b}}_{\pm}$ are the possible crossing states,
but $\gamma_2, \gamma_5, \delta$, and $\Delta_2$ (see Table \ref{tab:table2})
should be carefully evaluated to judge between them.

In Fig. 5(a), in addition to the anti-crossing
between $\rm{LL}0^{\rm{b}}_{-}$ and $-\rm{LL}3^{\rm{b}}_{-}$,
we also see anti-crossing between $\rm{LL}0^{\rm{m}}_{-}$ and $+\rm{LL}3^{\rm{b}}_{-}$.
This is in contrast with the case at $B = 2.8\rm{T}$ where $\rm{LL}0^{\rm{m}}_{-}$ and $+\rm{LL}3^{\rm{b}}_{-}$
degenerate but don't hybridize at $D = 0\rm{V/nm}$ (see Fig. 2(a)).
Under such a high $B$, $+\rm{LL}3^{\rm{b}}_{-}$ shifts to the high energy with $B$,
leading to the these LLs crossings at non-zero $D$.
This observation
indicates that both the trigonal warping term ($\gamma_3$) that hybridizes $\rm{LL}0^{\rm{b}}_{-}$ and $-\rm{LL}3^{\rm{b}}_{-}$ and the mirror symmetry breaking by $D$ that hybridizes $\rm{LL}0^{\rm{m}}_{-}$ and $\rm{LL}0^{\rm{b}}_{-}$ (H1) 
are required to form the anti-crossing structure.

In Fig. 5(d), we magnified the region of the anti-crossing 
between $\rm{LL}0^{\rm{m}}_{-}$ and $+\rm{LL}3^{\rm{b}}_{-}$ for $D < 0$.
While the process of the anti-crossing,
spin splitting of $\rm{LL}0^{\rm{m}}_{-}$ disappears and recovers again across the anti-crossing.
This observation is also consistent with anti-crossings, 
across which the LL index is preserved.

We also observe an effect of the hybridization (H3) and (H4) in Fig. 5(a).
It is the valley splitting
of $-\rm{LL}2^{\rm{b}}_{\pm}$ for $|D|>0.3\ \rm{V/nm}$.
$-\rm{LL}2^{\rm{b}}_{+}$ mainly hybridizes with $-\rm{LL}1^{\rm{m}}_{+}$ through the process (H4) while $-\rm{LL}2^{\rm{b}}_{-}$ mainly hybridizes with $-\rm{LL}2^{\rm{m}}_{-}$ through (H3).
Since the energy difference between $-\rm{LL}2^{\rm{b}}_{+}$ and $-\rm{LL}1^{\rm{m}}_{+}$ is smaller than that
between $-\rm{LL}2^{\rm{b}}_{-}$ and $-\rm{LL}2^{\rm{m}}_{-}$,
$-\rm{LL}2^{\rm{b}}_{+}$ -- $-\rm{LL}1^{\rm{m}}_{+}$ hybridization is stronger than 
$-\rm{LL}2^{\rm{b}}_{-}$ -- $-\rm{LL}2^{\rm{m}}_{-}$.
This results in valley splitting.

Through out the paper we neglected intervalley hybridization and
assumed that two valley states behave independently.
The actual device has rough edges because it was shaped by dry etching and therefore may experience
intervalley hybridization of the edge states.
However, this will not affect the bulk LLs which we mainly discussed above.
In the counter-propagating edge transport regime, electron and hole channels coexist 
and the channel number is influenced by hybridization of edge states.
In the $\nu = 2$ regime at $B = 5.6\ \rm{T}$ and $D = 0\ \rm{V/nm}$ (indicated by a green arrow in Fig. 5(a)),
usual two chiral edge transport ($R_{xx} \sim 0\ \Omega$ and $R_{xy} \sim \frac{1}{2}\frac{h}{e^2}$)
has been observed inspite of the expected counter-propagating edge transport
which consists of four electron channels and two hole channels.
This can involve the intervalley hybridization of the edge states by rough edges
where the total edge states are reduced to two chiral edge states.

In summary, we observed LL crossings under an electric field in ABA-stacked trilayer graphene,
and find that the overview of the LL evolution is consistent with the theory of hybridized 
MLG-like and BLG-like bands.
The observed LL evolution shows a variety of valley and orbital dependences. 
We find that it is explained by considering the hybridization between the multiple LLs sharing the same LL orbital and valley
and the hybridization between every third LL orbitals due to the trigonal warping term $\gamma_3$.
We also find pronounced agreement between our experimental observations and numerical calculation based on the
tight binding model \cite{Serbyn2013}.
We finally revealed the entire picture of the LL evolution with electric field in ABA-stacked trilayer graphene,
and showed a vatiery of quantum Hall phase can be achieved via the mirror symmetry breaking using electric field.
While preparing the manuscript, we found related works on the quantum Hall phases achieved in different regimes \cite{Campos2016, Stepanov2016}.